\documentclass[twocolumn,prb,nofootnoteinbib]{revtex4-1}

\usepackage{graphicx}
\usepackage{amsmath,amsthm,tikz,calc}
\usepackage{amssymb}
\usepackage{layout}
\usepackage{bm}   
\usepackage{bbold}
\usepackage{braket}
\usepackage{color}

\DeclareMathOperator{\sign}{sign}
\DeclareMathOperator{\linspan}{span}

\newcommand{\Ir}{\mathbb{Z}}
\newcommand{\Cx}{\mathbb{C}}
\newcommand\idty{\mathbb{1}} 
\newcommand{\cE}{\mathcal{E}}
\newcommand{\cF}{\mathcal{F}}
\newcommand{\cV}{\mathcal{V}}
\newcommand{\cH}{\mathcal{H}}
\newcommand{\hx}{\hat{x}}
\newcommand{\hy}{\hat{y}}
\newcommand{\be}{\begin{equation}}
\newcommand{\ee}{\end{equation}}
\newcommand{\bea}{\begin{eqnarray}}
\newcommand{\eea}{\end{eqnarray}}
\newcommand{\beann}{\begin{eqnarray*}}
\newcommand{\eeann}{\end{eqnarray*}}
\newcommand{\eq}[1]{(\ref{#1})}

\date{\today}

\begin{document}

\title{A dynamical Toric Code model with fusion and de-fusion}
\author{Bruno Nachtergaele$^{1}$, Nicholas E. Sherman$^{2}$}
\affiliation{
$^1$Department of Mathematics and Center for Quantum Mathematics and Physics, University of California, Davis, USA\\
$^2$Department of Physics, University of California, Berkeley, USA}
\begin{abstract}
We introduce a two-parameter family of perturbations of Kitaev's Toric Code Model in which the anyonic excitations acquire an interesting dynamics. We study the dynamics of this model in the space of states with electric and magnetic charge both equal to 1 and find that the model exhibits both bound states and scattering states in a suitable region of the parameters. The bound state is a Majorana fermion with a dispersion relation of Dirac cone type. For a certain range of model parameters, we find that these bound states disappear in a continuum of scattering states at a critical value of the total momentum. The scattering states describe separate electric and magnetic anyons, which in this model each have a $\sin k$ dispersion relation.
\end{abstract}
\maketitle

\section{Introduction}
% A good introduction... The goal here is to develop a perturbation for the Toric Code model (TCM) that introduces dynamics into the anyons of this model, while simultaneously preserving anyon number. We are ultimately interested in studying the dynamics of a single electric and a single magnetic anyon in this model. However, the anyon excitations of the TCM are always formed in pairs with the anyons connected by strings of Pauli matrices, and so we send one end of the string to infinity to look at a one particle excitation. Before discussing the model, we justify this process mathematically, and show that the physics is invariant under the way in which we take the endpoint to infinity.

There is increasing evidence that Majorana fermions and various types of anyons occur in the quantum many-body systems that exhibit 
topological order \cite{willett:2013,morampudi:2017,knapp:2019,takane:2019}. Toy models such as Kitaev's quantum double models \cite{kitaev:2003} and the Levin-Wen string-net 
models \cite{levin:2005} have been instrumental in showing that short-range lattice Hamiltonians in two dimensions can exhibit a rich variety of anyonic excitation spectra. Another important step was achieved by Bravyi, Hastings, and Michalakis \cite{bravyi:2010} and, independently, Klich \cite{klich:2010}, when they showed that the spectral gap above the ground state of these models is stable against sufficiently weak but otherwise arbitrary perturbations. In \cite{cha:2018a} it was shown that not only the gap, but the specific structure of anyon types and the associated superselection sectors too are stable in the same sense.

One of the virtues of the Toric Code Model (TCM) is that it is explicitly solvable and the structure of its eigenstates can be given explicitly and
fully understood. This is possible because it is a so-called `commuting Hamiltonian', meaning that all terms in the Hamiltonian \eq{HamTC}
commute and, 
hence, can be simultaneously diagonalized. It shares this property with the entire class of quantum double models introduced by 
Kitaev \cite{kitaev:2003} and many other models as well. The commuting property has a drawback, however, since it implies that the
model has no meaningful dynamics. The particle-like excitations, the anyons, are dispersionless (flat bands), and are 
therefore static. To explore the properties of anyons in 
a more realistic setting, we set out to modify the TCM by adding new finite-range interactions to Kitaev's Hamiltonian while at the 
same time aiming to preserve as many of its symmetries as possible. First and foremost, we want to preserve the anyon structure of the 
model. Equivalently, we want to preserve the superselection sectors, which are labeled by the charges corresponding to the quantum-double 
$\Ir_2\times\Ir_2$. We will label the four superselection sectors by $0$ (the vacuum sector), $\epsilon$ (odd electric and even magnetic charge), 
$\mu$ (even electric and odd magnetic charge), and $\epsilon\mu$ (electric and magnetic charges both odd). The TCM does not only preserve the 
topological charge, given by the parities of the electric and magnetic charge but, via the energy, in fact preserves the integer values of both types 
of charges (particle number conservation). By respecting these conservation laws we will guarantee that the vacuum state itself is left invariant 
under the perturbations we consider and it also implies the existence of an invariant subspace of states with one electric and one magnetic anyon 
present. In this paper, our primary interest is the spectrum of the Dynamical Toric Code Model (DTCM) that we introduce below, in that subspace.

\section{The space of states with unit topological charge}

Kitaev's TCM is a two-dimensional quantum spin model that is commonly defined on the regular square lattice $\Ir^2$. It has a duality symmetry that is most directly seen if we use the \emph{edges} of the square lattice to label the spins. Let $\cV$, $\cE$, $\cF$, denote the set of vertices, the set of edges, and the set of faces, respectively, of the regular square lattice. For each $v\in\cV$ and $f\in \cF$, define
\be
A_v = \prod_{e, v\in e} \sigma^1_e,\quad B_f \prod_{e, e\in f}\sigma^3_e.
\ee
These are often referred to as the star and plaquette operators of the toric code. In terms of these, the TCM is defined by the Hamiltonian
\be
H^{TC} = \sum_{v\in \cV} (\idty-A_v) + \sum_{f\in\cF} (\idty- B_f).
\label{HamTC}\ee 

For our purposes, the infinite lattice setting is best suited. First, the difference between bound states and scattering states is a clear mathematical distinction for the infinite system. Second, there are well-defined sectors of the model with single anyon excitations, while in finite volume such excitations are always 
created in pairs starting from the vacuum. By considering the infinite lattice, we can consider the limit where one of the anyons of a pair is taken
to infinity. This leads to a simple structure of superselection sectors \cite{naaijkens:2011, cha:2018} which has recently been shown to be stable under perturbations of the Hamiltonian \cite{cha:2018a}. Let us recall the main features of the TCM on the infinite lattice in some detail. 

As shown in \cite{alicki:2007}, the TCM on infinite lattice $\Ir^2$ has a unique frustration free ground state, meaning there is a unique state fully determined by the
vanishing of all terms in the Hamiltonian:
$$
\omega(\idty-A_v) = \omega(\idty-B_f)=0, \mbox{ for all } v\in\cV, f\in\cF.
$$
Here, the state is represented by its expectation functional $\omega$. It will be useful to use a representation of this state as a 
unit vector $\Omega$ in the vacuum Hilbert space $\cH_0$:
\be
\omega(A) = \langle \Omega, \pi(A) \Omega\rangle,
\label{GNS}\ee
where $A$ is an arbitrary observable involving a finite set of spins, and $\pi$ is a representation of the observables acting 
on the Hilbert space $\cH_0$. $\cH_0$ itself represents all excitations of the model that can be created by such an observable 
acting on the vacuum state. This includes pairs of excitations created by finite string operators (see below). The representation 
\eq{GNS} is known as the Gelfand-Naimark-Segal (GNS) representation of the vacuum state \cite{bratteli:1987}. The vacuum vector $\Omega$ is 
characterized by the property that it is an eigenvector with eigenvalue 1 of each of the operators $A_v$ and $B_f$:
$$
\pi(A_v) \Omega = \pi(B_f)\Omega = \Omega, \quad v\in \cV, f\in \cF.
$$

Energy eigenstates in the vacuum sector are created by the action of the so-called string operators. These are associated with paths over
edges of either the lattice or the dual lattice as follows. Let $\gamma$ and $\tilde\gamma$ be a finite path in the lattice and the dual
lattice, respectively, and define
$$
F^\epsilon_\gamma = \prod_{e\in\gamma} \sigma^3_e,\quad  F^\mu_{\tilde\gamma} = \prod_{e\in\tilde\gamma} \sigma^1_e.
$$
Here, we have used the fact that the edges of the lattice and the edges of the dual lattice are in one-to-one correspondence. Hence, we can
use the same notation for them. Moreover, each spin is associated with exactly one edge of the lattice and one edge of the dual lattice.
Using the commutation relations of the Pauli matrices (which are preserved by any representation $\pi$), it is straightforward
to check that the vectors
$$
\pi(F^\epsilon_\gamma)\Omega, \mbox{ and } \pi(F^\mu_{\tilde\gamma} )\Omega
$$
are eigenstates of the Hamiltonian and that these vectors only depend on the end-points of the paths. In particular, closed paths leave 
$\Omega$ invariant, and each string operator corresponding to a finite open path generates an eigenstate with energy $4$, corresponding
to an eigenvalue $-1$ for the two terms $A_v$ and $A_{v^\prime}$ where $v$ and $v^\prime$ are the end points of the path $\gamma$ 
or $B_f$ and $B_{f^\prime}$, with $f$ and $f^\prime$ the two end faces of the dual path $\tilde\gamma$.
Due to the invariance of $\Omega$ under closed string operators, which is an expression of the gauge invariance of the model,
the eigenstates created by a single open string operator are consistently defined, including their phase. This changes when we consider
the states created by the combined action of a string operator and a dual string operator. The two anti-commute when the strings intersect an
odd number of times and commute otherwise. The vectors created from the vacuum by the action of both a string operator and a 
dual string operator on $\Omega$ may therefore differ by a minus sign depending on the choice of paths connecting a given pair of end points.
This does not affect expectation values of any observable in the excited states, but is relevant when calculating matrix elements of 
operators relating different excited states.
 
The expectation functionals for excited states with just one anyon, of which energy equals 2, can be defined unambiguously by considering 
a sequence of paths with one end point held fixed while the other end point is taken to infinity. In such a limit, the expectation value of any 
observable depends only on the location of the end point that is held fixed. This is easily verified using the invariance of the vacuum under the
action of closed-string operators.

It will be convenient to express the single anyon states as a modified representation of the algebra of observables. Note that
the operators $F^\epsilon_\gamma$ and $F^\mu_{\tilde\gamma}$ are self-adjoint unitaries that square to $\idty$. Therefore, conjugation 
with these operators defines a family of automorphisms of the algebra of observables. In order to set up a convention for a Hilbert space
of excitations as vectors states, we now fix a convention for how the endpoints of the paths are taken to infinity.
%When acting on $\Omega$, these operators create pairs of excitations localized at the endpoints of the corresponding path. These two excitations are conjugate to each other, meaning they
%annihilate each other when they meet at the same edge, as is the case in a closed loop. Therefore, $F^\epsilon_\gamma$ and $F^\mu_{\tilde\gamma}$ for \emph{closed} $\gamma$ and $\tilde\gamma$ leave $\Omega$ invariant. This property implies that the
%vectors  $F^\epsilon_\gamma\Omega$ and $F^\mu_{\tilde\gamma}\Omega$ only depend on the \emph{end points} of $\gamma$ 
%and $\tilde\gamma$.
We introduce families of paths $\gamma_v(n)$ and $\tilde\gamma_f(n)$ which
start at a vertex $v$ and face $f$, respectively, and stretch over $n$ edges and dual edges in the negative vertical direction, from
$v$ to $v-n \hat y$ and similarly for $f$ in the dual lattice. We can now define the following automorphisms of the observable algebra as
limits along these paths:
\bea
\alpha_v (A) &=& \lim_{n\to\infty} F^\epsilon_{\gamma_v(n)} A F^\epsilon_{\gamma_v(n)} \\
\alpha_f (A) &=& \lim_{n\to\infty} F^\mu_{\tilde\gamma_f(n)} A F^\mu_{\tilde\gamma_f(n)} 
\eea
It is easy to see that all these automorphisms commute and that the families $\{\alpha_v \mid v\in \cV\}$
and $\{\alpha_f \mid f\in \cF\}$ are covariant with respect to the automorphisms representing lattice translations: for all
$x\in \Ir^2$ one has
\beann
T_x \circ \alpha_{v} &=& \alpha_{v+x} \circ T_x \\
T_x \circ \alpha_{f} &=& \alpha_{f+x} \circ T_x,
\eeann
where $v+x$ and $f+x$ denote the translated vertex and face in the lattice. We are interested in the family of 
states 
$$
A\mapsto \omega_{v,f}(A):=\langle \Omega, \pi \circ\alpha_v\circ\alpha_f(A)\Omega\rangle.
$$
It is easy to see that the representations $\pi \circ\alpha_v\circ\alpha_f$ are all unitarily equivalent.
Therefore, the states $\omega_{v,f}$ can all be represented as vector states in one and the same Hilbert space.
This Hilbert space, however, is distinct from the one that contains the vacuum vector. It is an in-equivalent 
superselection sector, corresponding to different values of the topological charges.

Furthermore, if $(v,f)\neq (v^\prime,f^\prime)$, there is $X\in\{A_v,B_f\}$ such that the expectations $\omega_{v,f}(X)$
$\omega_{v^\prime,f^\prime}(X)$ are given by distinct eigenvalues of $X$. This implies that 
the vector states are mutually orthogonal. As a result, there is a natural identification of the
states with a single electric and magnetic excitation with $\ell^2(\cV \times \cF)
\cong\linspan \{ \psi(v,f) | v\in\cV, f\in\cF\}=: \cH^{\epsilon\mu}$ on which the translations act by 
unitaries $U_x, x\in \Ir^2$, in the canonical way. 

Note that we can reason in exactly the same way to construct Hilbert spaces $\cH^{\epsilon}$ and $\cH^{\mu}$ of 
states with a single electric and magnetic excitation, respectively. Each of the three Hilbert spaces $\cH^{\epsilon}, \cH^{\mu}$,
and $\cH^{\epsilon\mu}$ belong to its own superselection sector distinct from one another and from the vacuum 
sector \cite{naaijkens:2011}. We will use these spaces to motivate the perturbation terms we introduce to define the 
Dynamical Toric Code Model (DTCM) in the next section. We have that $\cH^{\epsilon}\cong \ell^2(\cV)$, $\cH^{\mu}\cong \ell^2(\cF)$,
and $\cH^{\epsilon\mu}\cong \ell^2(\cV\times \cF)$. We will use the natural orthonormal bases of anyon excitations 
$\{\ket{\epsilon_v}\}_{v\in\cV}$, $\{\ket{\mu_f}\}_{f\in\cF}$, and $\{\ket{\epsilon_v,\mu_f}\}_{v\in\cV, f\in\cF}$ for these spaces.

\section{A Dynamical Toric Code model}

Since the static anyon excitations $\ket{\epsilon_v}$ and $\ket{\mu_f}$, loosely speaking, correspond to the action of a half-infinite string operator
starting at $v$ and $f$, hopping terms that move the excitation should move the end point of the corresponding string operator. This amounts to 
the action of a Pauli matrix at a neighboring spin (extending the path) or at the end point itself, which shrinks the path by one unit. The action of the
Pauli matrices by themselves at a generic location, however, would create a pair of additional excitations and create a state of energy equal to 
$6$ and orthogonal
to the space of single anyon excitations. In order to achieve our goal of leaving the spaces $\cH^{\epsilon}$ and $\cH^{\mu}$ invariant 
under the action of the hopping terms, we use the operators $A_v$ and $B_f$ to detect the location of the excitation. It turns out that 
hopping matrix elements are imaginary and have the correct sign with respect to a reference orientation, which by the gauge 
symmetry one can freely chose. There is complex, non translation-invariant gauge transformation that makes the hopping matrix elements real, 
but working in that gauge would not offer any advantages.
A simple translation-invariant orientation is the following: let all horizontal edges point to the right and all vertical edges point up.
In terms of this orientation we define a sign function on the pairs $(v,e)\in\cV\times\cE, v\in e$ as follows:
Define $s(v,e)=1$, if $e$ is outgoing with respect to $v$, and $s(v,e)=-1$, if $e$ is incoming
with respect to $v$. The sign of a face $f$ with respect to an edge $e$, denoted by $s(f,e)$ is defined consistent with
the duality of faces to vertices: $s(f,e)=1$ if $f$ is below or to the left of $e$ and $s(f,e)=-1$ if $f$ is above or to the right of $e$.

The hopping terms that satisfy our criteria are then
\begin{equation}
    H^{\epsilon} := i\sum_e\sigma^3_e\sum_{v\in e}s(v,e)A_v.
\end{equation}
and 
%\begin{equation}
%    H^{\epsilon}\ket \Omega = 0
%\end{equation}
%and so since these terms annihilate the vacuum, they also can't hop anyons in such away to annihilate each other. 
\begin{equation}
    H^{\mu} := i\sum_e\sigma^1_e\sum_{f\ni e}s(f,e)B_f
\end{equation}

To find the spectrum, we first find the matrix elements of $H^\epsilon$ and $H^\mu$, restricted to the invariant subspaces $\cH^\epsilon$
and $\cH^\mu$, with respect to the orthonormal bases $\{\ket{\epsilon_v}\}_{v\in\cV}$ and $\{\ket{\mu_f}\}_{f\in\cF}$, respectively. 
To do this, we identify the vertex set $\cV$ with a copy of the integer lattice $\Ir^2$ and define $S$ to be the set of unit lattice vectors: $S=\{\hat x,-\hat x,\hat y,-\hat y\}$. 
Then, the matrix elements of $H^{\epsilon}$ restricted to $\mathcal{H}^{\epsilon}$ are given by
\begin{equation}
    \bra{\epsilon_{v'}}H^{\epsilon}\ket{\epsilon_v} = 2i\sum_{r\in S}(r\cdot\hat x + r\cdot \hat y)\delta_{v',v+r}
\end{equation}
The spectrum and the dispersion relation are then easily found by Fourier transformation:
\begin{equation}
    E(k) = 4\sin(k_x) + 4\sin(k_y), k_x,k_y \in [0,2\pi).
\end{equation}
The magnetic anyons described by $H^\mu$ have the same spectrum.

%In an identical way, we can define a one magnetic anyon state
%\begin{equation}
%    \ket{\mu_f} = F^{\mu}_{\tilde\gamma_f}\ket{\Omega}
%\end{equation}
%
%and we fine the same spectrum for $H^{\mu}$ when restricted to $\mathcal{H}^{\mu}$.

Since we are interested in seeing the dynamical properties of interacting anyons and specifically the merging of an electric and 
a magnetic excitation into an excitation created by what is called a ribbon operator \cite{bombin:2008}, we also need to consider 
adding terms to the Hamiltonian that describe electric-magnetic interactions.
Formally, we again have an orthonormal
basis of $\cH^{\epsilon\mu}$ with one electric anyon at the vertex $v$ and one magnetic anyon at the face $f$ created
by a pair of string operators, with the same convention of paths and dual paths extending to $\infty$ in the negative $\hy$ direction:
\begin{equation}
    \ket{\epsilon_v\mu_f} = F^{\epsilon}_{\gamma_v}F^{\mu}_{\tilde\gamma_f}\ket\Omega.
\end{equation}
If $v$ and $f$ are such that there is an edge $e$ satisfying $v\in e$, $e\in f$ ($v$ and $f$ are next to each other) then this represents a fused $\epsilon\mu$ state that we denote by $\ket{\epsilon\mu_s}$, where $s$ is a `site' determined by a  pair $(v,f)$, where $v$ is a vertex belonging to the face $f$.
The ribbon states correspond to the pairs $(v,f)$ with $v\in f$. Using the same principles as for the single $\epsilon$ and $\mu$ hopping 
terms we constructed a hopping term that leaves the subspace of ribbon states invariant:
%\begin{equation}
%    H^{\epsilon} + H^{\mu}
%\end{equation}
%however, this combination will break apart an $\epsilon\mu$ particle into two seperate particles. To prevent this, we must add a term ensuring that hopping either the $\epsilon$ or $\mu$ portion of the particle couples to a new $\epsilon\mu$ state. If we hop the electric anyon, it must hop over an edge $e$ that has an magnetic anyon on a face $f\ni e$. Similarly, if we hop a magnetic anyon, it must hop over an edge $e$ that has an electric anyon at the vertex $v\in e$. We enforce this condition with the term
\begin{align}
    H^{\epsilon\mu}=&i\sum_{e}\sigma^3_e\sum_{v\in e} s(v,e) A_v\sum_{f\ni e} (\idty - B_f)\\
 + &i\sum_{e}\sigma^1_e\sum_{f\ni e} s(f,e) B_f\sum_{v\in e}(\idty - A_v).
\end{align}

Note that, individually, the terms $H^\epsilon, H^\mu$, and $H^{\epsilon\mu}$ leave the sectors with one $\epsilon$, one $\mu$, and one $\epsilon$ plus one $\mu$ excitation invariant. Ribbon states, however, may be broken up by  $H^\epsilon$ and $H^\mu$.
Using these three terms we define the Hamiltonian of the Dynamical Toric Code model (DTCM) as follows:
\be
H^{\rm DTC} = H^{\rm TC} + \lambda_\epsilon H^\epsilon + \lambda_\mu H^\mu + \rho H^{\epsilon\mu}.
\label{HamDTCM}\ee

\section{Symmetries}

Before analyzing the DTCM, let us observe important symmetries of $H^{\rm TC}$ and $H^{\rm DTC}$. In addition to the translation and 4-fold rotation symmetry of the lattice, the TCM also has lattice inversion (aka parity), time reversal, spin-flip, charge conjugation, chirality, and duality symmetries, which we now discuss.

Lattice inversion or parity symmetry stems from the lattice invariance under reflection through the origin: $(x_1,x_2)\to (-x_1,-x_2)$. 
The terms appearing in the perturbations $H^\epsilon, H^\mu$, and $H^{\epsilon\mu}$ all anti-commute with inversion due to the presence 
of the signs $s(v,e)$ and $s(f,e)$. The vacuum state is invariant under parity. Therefore the symmetry is represented by a unitary operator $P$
in the GNS representation (as is also the case in finite volume, of course).

Since $P$ commutes with $H^{\rm TC}$ but anti-commutes with the perturbation terms in $H^{\rm DTC}$,
the perturbed spectrum in the single anyon sectors is symmetric around the unperturbed excitation energy, as is illustrated in Figure \ref{fig:energies}.

Time reversal, an anti-unitary $T$ with $T^2=\idty$, implemented by complex conjugation combined either spin flip ($T\sigma^i = -\sigma^i T, i=1,2,3$) or parity. 
We have $[H^{\rm DTC},T]=0$.
This has an important implication for the spectrum of $H^{\epsilon\mu}$ restricted to $\mathcal{H}^{\epsilon\mu}$, as the $\epsilon\mu$ particle has fermion self-statistics. 
This means in this sector that Kramer's degeneracy implies an even degeneracy for all values of the spectrum.

Charge conjugation is an anti-unitary $C$ with $C^2=\idty$, implemented by complex conjugation by itself. We have $\{H^{\rm DTC} - H^{\rm TC},C\} = 0$ and $[H^{\rm TC},C]=0$.

Chirality is described by a unitary $S$ with $S^2 = \idty$, implemented by either spin flip or parity. We have $\{H^{\rm DTC}-H^{\rm TC},S\}=0$ and $[H^{\rm TC},S]=0$.

The $\Ir_2\times\Ir_2$ symmetry given by the spin rotations by $\pi$ are implemented by the conjugation with the Pauli matrices. These `spin flip' symmetries are broken in $H^{\rm DTC}$. 

Duality symmetry, which interchanges the lattice and the dual lattice, is implemented by a local unitary operator taking the form
\begin{equation}
    D = \frac{1}{\sqrt{2}}\left(
    \begin{matrix}
    1 & 1 \\
    1 & -1
    \end{matrix}\right)
\end{equation}
D has the following properties
\begin{align}
    D^2 &= \idty\\
    D\sigma^1D &= \sigma^3\\ 
    D\sigma^3 D &= \sigma^1 \\
    D\sigma^2D &= -\sigma^2
    % D &= e^{-i\frac{\pi}{2}\sigma^1}e^{-i\frac{\pi}{4}\sigma^2}
    % D&:\left\{\begin{matrix}
    % \sigma^1\rightarrow \sigma^3 \\
    % \sigma^2 \rightarrow -\sigma^2 \\
    % \sigma^3 \rightarrow \sigma^1
    % \end{matrix}\right.
\end{align}
Duality interchanges $\epsilon$ and $\mu$ excitations:
\begin{align}
    [D,H^{\rm TC}] &= [D,H^{\epsilon\mu}] = 0 \\
    DH^{\epsilon}D &= H^{\mu}
\end{align}
Therefore, if $\lambda_{\mu} = \lambda_{\epsilon} := \lambda$, we have that $[D,H^{\rm DTC}] = 0$.
%
%\nick{We may want to adjust this section a bit to be more in line with the traditions in the periodic table of topological invariants. We have the following in their language
%
%Time reversal, an anti-unitary $T$ with $T^2=1$, implemented by complex cunjugation times either spin flip or parity. We have $[H^{\rm DTC},T]=0$.
%
%Charge conjugation, an anti-unitary $C$ with $C^2=1$, implemented by just complex conjugation. We have $\{H^{\rm DTC} - H^{\rm TC},C\} = 0$ and $[H^{\rm TC},C]=0$.
%
%Chirality, a unitary $S$ with $S^2 = 1$, implemented by either spin flip or parity. We have $\{H^{\rm DTC}-H^{\rm TC},S\}=0$ and $[H^{\rm TC},S]=0$.
%
%We also have Duality. Note, with these definitions, we also have $CP = T$, and $T^2=1$, and so we have $CPT=1$ is a trivial symmetry, although I am not sure this is meaningful, but maybe explains why we have a Dirac-cone? Not sure
%}
%\bxn{I implemented most of your suggestion. See above. I am not sure CPT is worth mentioning here. I think not.}

\section{The Hamiltonian in the $\epsilon\mu$ sector}

As mentioned above, the Hilbert space of single-anyon excitations in the $\epsilon,\mu$ and $\epsilon\mu$ sector are individually 
left invariant. Therefore, the dispersion relation for each anyon type is well-defined and is easily computed as long as one takes
care to define a suitable basis in the appropriate Hilbert space. The Hilbert space of states with exactly one electric and one magnetic anyon,
$\cH^{\epsilon\mu}$, is a subspace of the $\epsilon\mu$ sector and is invariant for the Hamiltonian $H^{DTCM}$ defined in
\eq{HamDTCM}. To calculate its spectrum we will find its matrix elements with respect to the basis $\{\ket{\epsilon_v \mu_f}\}_{v\in\cV,f\in\cF}$.

Since the vacuum state is also the ground state of the DTCM, at least for $\lambda$ and $\rho$ not too large
\cite{bravyi:2010} (and generally is a stationary state), the dynamics of the states $\ket{\epsilon_v \mu_f}$ can be 
studied in terms of commutation properties of the Hamiltonian with the operators $F^\epsilon_{\gamma_v(n)}$ and 
$F^\mu_{\tilde\gamma_f(n)} $ and the property that the vacuum state is invariant under closed loop operators.
%Explicitly:
%\beann
%&&\pi \circ\alpha_v\circ\alpha_f\circ\tau^{\epsilon\mu}_t (A)\\
%&&=\lim_n U^{\epsilon\mu}(t)^* \pi ( \tau^{\epsilon\mu}_{-t} (F^\epsilon_{\gamma_v(n)}  F^\mu_{\tilde\gamma_f(n)} )A
% \tau^{\epsilon\mu}_{-t} (F^\epsilon_{\gamma_v(n)}  F^\mu_{\tilde\gamma_f(n)} )) U^{\epsilon\mu}(t).
%\eeann
It follows that we can analyze the dynamics of the states $\ket{\epsilon_v \mu_f}$ in terms of a Hamiltonian $h^{\epsilon\mu}$
on $\ell^2(\cV\times\cF)$, which is unitarily equivalent to the invariant subspace $\cH^{\epsilon\mu}$. All we need
to do is calculate the matrix of $H^{DTCM}$ with respect to the basis states $\ket{\epsilon_v \mu_f}$.
%given by
%\beann
%&&\lim_n\sum_{v^\prime,f^\prime} h^{\epsilon\mu}_{v^\prime f^\prime, v  f} 
% \pi(F^\epsilon_{\gamma_{v^\prime}(n)}  F^\mu_{\tilde\gamma_{f^\prime}(n)})\Omega\\
%&&= - \lim_\Lambda \lim_n \pi([H^{\epsilon\mu}_\Lambda, F^\epsilon_{\gamma_v(n)}  F^\mu_{\tilde\gamma_f(n)}])\Omega
%\eeann

The term $H^{TC}$ acts as a constant (=4) on the subspace $\cH^{\epsilon\mu}$. 
The dynamics of the DTCM restricted to $\cH^{\epsilon\mu}$ is therefore solely due to the terms 
$H^\epsilon, H^\mu$, and $H^{\epsilon\mu}$. It is also clear from its definition that 
$h^{\epsilon\mu}$ commutes with the lattice translations acting on $\cH^{\epsilon\mu}$.
After taking the Fourier transform with respect to the `center of mass' coordinates $X$,
and writing $(v,f) = (X-d, X+d)$, we obtain a useful expression for $h^{\epsilon\mu}$
on the subspace of total quasi-momentum $K$. It is convenient to consider the even and 
odd square sublattices of $\Ir^2$ to label the vertices $\cV$ and the faces $\cF$, respectively.
This implies that the relative coordinates $d$ are to be taken in $(2\Ir +1)^2$. We will use the notation
$d=d_1 \hx + d_2 \hy$, where $\hx =(1,0), \hy=(0,1)$. We also define the function $\theta $ on the
odd integers by $\theta (d) = (1-\sign(d))/2$.

%
%The dispersion relations of $H_\epsilon $ and 
%$H_\mu$ are given by
%$$
%\omega(k) = 4(\sin k_1 +\sin k_2).
%$$
%$H_{\epsilon\mu}$  also leaves the `ribbon subspace invariant', which we define as the space of excitations of type $\epsilon\mu$ residing on
%a site $(v,f)$, defined as a face $f$ and one of the vertices $v$ belonging to $f$ (see, e.g., \cite{bombin:2010}. The ribbon excitations are
%Majorana fermions. Their dispersion relation exhibits a Dirac cone at $k=0$, and is given by:
%$$
%\omega_r(k) = \pm8\sqrt{1 -(\cos k_1 +\cos k_2)/2}.
%$$
%
%Since the subspace of states with one anyon of each type
%is an invariant subspace of the dynamics generated by
%$$
%H = H^{TC} + \lambda ... + \rho ...
%$$
%This subspace isomorphic to $\ell^2(\Ir^2)\otimes \ell^2(\Ir^2)$
%The reduction of $H$ to this subspace, of course, commutes
%with translations and therefore total momentum is conserved.
%Performing the Fourier transform block-diagonalizes the 
%Hamiltonian, where each block is labeled by the total 
%momentum $K\in [0,\pi]^2$ \bxn{check conventions}.
%
%For ease of the book keeping, we will describe the relative coordinates
%of the two anyons by $r\in (2\Ir+1)^2)$, i.e., a two-component vector
%with odd integer components. 

A careful calculation yields the following matrix elements of $h$ in the subspace of states 
with total quasi-momentum $K=(K_1,K_2)$: for $d',d\in (2\Ir +1)^2$ we have 
\beann
(h_K)_{d',d} &=& -2\lambda H^0_{d',d}(K)  -2\rho H^i_{d',d}(K) \\
(H^0_K)_{d',d} &=& \sin 2K_1 [ \delta_{d',d-2\hx}(-1+2\delta_{d_1,1}\theta(d_2))\\
&& +\delta_{d',d+2\hx}(-1+2\delta_{d_1,-1}\theta(d_2))]\\
&& - \sin 2K_2 [ \delta_{d',d-2\hy}  +\delta_{d',d+2\hy}]\\
(H^i_K)_{d',d} &=& \sin 2K_1 [ \delta_{d',d-2\hx}\delta_{d_1,1} (\delta_{d_2,1}-\delta_{d_2,-1})\\
&& + \delta_{d',d+2\hx}\delta_{d_1,-1} (\delta_{d_2,1}-\delta_{d_2,-1})]\\
&& -\sin 2K_2 [ \delta_{d',d-2\hy}\delta_{d_2,1} (\delta_{d_1,1}+\delta_{d_1,-1})\\
&& + \delta_{d',d+2\hy}\delta_{d_2,-1} (\delta_{d_1,1}+\delta_{d_1,-1})].
\eeann
% \nick{I Changed the matrix elements, and $H_K$ below to what I believe to be correct}
To study the spectrum, we will consider the Hamiltonian on $\cH^{\epsilon\mu}$
as the bounded self-adjoint operators on $\ell^2((2\Ir+1)^2)$ of the following
form: 
\be
H_K = 4\idty - 2\lambda H^0_K - 2\rho H^i_K.
\label{H_K}\ee
To analyze $H_K$ it will be convenient to regard it 
as an operator on $\ell^2(2\Ir+1)\otimes \ell^2(2\Ir+1)$,
with the two factors corresponding to the $x$ and $y$ 
components of $r$. We then find
\beann
H^0_K &=& \sin (2K_y) (\idty \otimes \Delta) + \\
&&\sin(2K_x) [\Delta\otimes (\idty - \Theta) +(2\Theta-\idty)\Delta (2\Theta-\idty)\otimes \Theta],
\eeann
where $\Delta$ is the discrete Laplace operator with zeros on the diagonal
and $\Theta$ is the diagonal operator with $\Theta_{ii} = 1$
for $i>0$ and $\Theta_{ii} = 0$ for $i<0$, $i\in 2\Ir+1$.
 
To describe $H^i_K$, consider the ordered set $S$ of 4 nearest neighbors
in $(2\Ir +1)^2$ given  by $S=\{(1,1), (1,-1), (-1,1), (-1,-1)\}$, and denote
by $P_S$ the orthogonal projection onto the states in $\ell^2((2\Ir +1)^2)$
that have zero components outside of $S$. Then, $H^i_K =
P_S H^i_K P_S$ and the $4\times 4$ block corresponding 
to $S$ is given by
\be
\begin{pmatrix}
0& -\sin(2K_y) & \sin(2K_x)  & 0 \\
-\sin(2K_y)&0& 0 & -\sin(2K_x) \\
\sin(2K_x)&0& 0 & -\sin(2K_y) \\
0& -\sin(2K_x) & -\sin(2K_y)  & 0 
\end{pmatrix}
\ee

Since $H^i_K$ is of finite rank, the essential spectrum of $H_K$ is
the spectrum of $4\idty - 2\lambda H_K^0$, which is purely continuous.
This means that the two spectra can only differ by one or more eigenvalues.

\section{The spectrum in the $\epsilon\mu$ sector}\label{sec:spectrum}

We start with finding the eigenvalues of $H^i_K$. Since, the only
non-zero matrix elements are in a $4\times 4$ block, $0$ is an infinitely
degenerate eigenvalue in the thermodynamic limit and, in addition, we have
the eigenvalues of the block, which can be compactly written as
\be
P_S H^i_K P_S
 = \sin(2K_x) (\sigma^1\otimes \sigma^3) -\sin(2K_y) (\idty\otimes\sigma^1).
\ee
Therefore, the non-zero eigenvalues of $H^i_K$ are easily seen to be 
$$
\pm\sqrt{\sin^2(2K_x) + \sin^2(2K_y)},
$$
which are both doubly degenerate, in agreement with the Nielsen-Ninomiya Theorem \cite{nielsen:1981,friedan:1982}. 
This is a typical Dirac cone and, in contrast with some claims in the literature, space-time inversion symmetry ($TP$)
is not required for this feature \cite{wang:2015}.
% This is a typical Dirac cone,  In
% contrast with some claims in the literature, a hexagonal structure 
% is not required for this feature \cite{wang:2015}. 
% \nick{This reference doesn't claim that hexagonal is required, just that it is favorable for Dirac cones, and the square lattice is not. It does however state that most, if not all, Dirac cones are induced by space-time inversion symmetry (time-reversal composed with parity) which we don't have}

The norm of $H^0_K$ is easily seen to be given by
\be
\Vert H^0_K \Vert =  2|\sin(2K_x)| + 2|\sin (2K_y)|.
\label{normHaK}\ee
To study the spectrum of $H^0_K$ it is convenient to rewrite this 
operator as
\beann
H^0_K &=&  \sin(2K_x) (\Delta\otimes \idty)  +   \sin(2K_y)  (\idty \otimes \Delta)\\ 
&&+  \sin(2K_x) \left([(2\Theta-\idty)\Delta (2\Theta-\idty) -\Delta ]\otimes \Theta\right).
\eeann
Note that the operator between square brackets has only two non-vanishing 
matrix elements in the canonical basis of $\ell^2(2\Ir +1)$. Using this fact
and the standard plane waves as approximate eigenvectors, we
then easily find that the spectrum is given by the values 
$$
E(K,k) = 2\sin 2K_x \cos 2k_x + 2\sin 2K_y \cos 2k_y,
$$
for $K,k\in [-\pi/4,\pi/4]$. 
There is no indication of the existence of bound states (eigenvectors
in $\ell^2$) and we therefore expect that the spectrum is purely absolutely 
continuous.

Using the norm \eq{normHaK}, we see that when
\be
\frac{|\lambda|}{|\rho|} < \frac{1}{4}\frac{ \sqrt{\sin^2(2K_x) + \sin^2(2K_y)}}{|\sin(2K_x)| + |\sin(2K_y)|}
\label{perturbative_regime}\ee
$H_K$ will have two eigenvalues $E_\pm(K,\lambda)$. The condition \eq{perturbative_regime}
is satisfied for all $K$ if 
$$
\frac{|\lambda|}{|\rho|} \leq \frac{1}{4\sqrt{2}}.
$$
For larger ratios, it is possible that the eigenvalues persist only for a restricted range 
of values of the total momentum.

In addition to the eigenvalues discussed above, which represent bound states of the 
two anyons, for $\lambda \neq 0$, there is also a band of scattering states in which
the two anyons are unbound.

%From numerical results discussed later, the most strongly bound $K$ values reside at the edge of the Brillouin zone boundary with $K_x$ or $K_y$ equal to 0. 
For values of $K$ at the Brillouin zone boundary, i.e., $H=(K_x,0)$ or $K=(0,K_y)$, the Hamiltonian $H_K$ becomes essentially separable and is equivalent to a family
of one-dimensional Hamiltonians. For these $K$ values, we can find the exact range of the parameters that produce a bound state in the spectrum by a transfer matrix 
analysis. To see this, we rewrite $H_K$ of \eq{H_K} as follows. 

For simplicity, we focus our attention to the case $K = (0,K_y)$. Recall that we regard $H_K$ as an operator on $\ell^2(2\Ir +1)\otimes \ell^2(2\Ir +1)$. If we define the
rank-2 projection $P^\pm = \ket{1}\bra{1} + \ket{-1}\bra{-1} $ on $\ell^2(2\Ir +1)$, we see $P_S = P^\pm\otimes P^\pm$. We then have
\beann
&&H_{(0,K_y)} =4\idty - 2\lambda \sin (2K_y) G\\
&&G = \idty \otimes \Delta -\frac{\rho}{\lambda}  P^\pm\otimes ( \ket{1}\!\!\bra{-1} + \ket{-1}\!\!\bra{1} ).
\eeann
By using the basis $\ket{x}, x\in 2\Ir^2 +1$, for the first tensor factor, $G$ can be further decomposed as follows:
\beann
G&=& (\idty-P^\pm)\otimes \Delta \\
&& + P^\pm\otimes [\Delta -\frac{\rho}{\lambda}  ( \ket{1}\!\!\bra{-1} + \ket{-1}\!\!\bra{1} )].
\eeann
This is an orthogonal decomposition showing that the spectrum of $G$ is the union
of the spectra of $\Delta$ and the spectrum of the operator 
$$
 \Delta' = \Delta -\frac{\rho}{\lambda} ( \ket{1}\!\!\bra{-1} + \ket{-1}\!\!\bra{1} ).
$$
Concretely, $\Delta'$ is a bi-infinite tri-diagonal matrix of the following form:
$$
\Delta'=
\left[\begin{array}{cccccc}
\ddots&\ddots&&&&\\
\ddots&0&-1&&&\\
&-1&0&-1-\rho/\lambda&&\\
&&-1-\rho/\lambda&0&-1&\\
&&&-1&0&\ddots\\
&&&&\ddots&\ddots
\end{array}
\right]
$$
We want to find
\begin{equation}\label{eig}
    \Delta'\ket{\psi} = E\ket{\psi},
\end{equation}
where the spectral value $E$ corresponds to a bound state if we have a non-zero solution $\ket{\psi}\in l^2(2\mathbb{Z} + 1)$. 
Scattering states correspond to $E$ values with $\ket{\psi}$ that are 
not square-summable.

Equation (\ref{eig}) gives a system of equations for the components $\psi(n)$ of $\ket{\psi}$ as follows
\begin{align}
    \left(\begin{matrix}
    \psi(2n-1) \\
    \psi(2n+1)
    \end{matrix}\right) &= T
    \left(\begin{matrix}
    \psi(2n-3) \\
    \psi(2n-1)
    \end{matrix}\right),\quad n\ge 2 \\
    \left(\begin{matrix}
    \psi(2n-1) \\
    \psi(2n+1)
    \end{matrix}\right) &= S
    \left(\begin{matrix}
    \psi(2n+1) \\
    \psi(2n+3)
    \end{matrix}\right),\quad n\le -2\\
    \left(\begin{matrix}
    \psi(1) \\
    \psi(3)
    \end{matrix}\right) &= A
    \left(\begin{matrix}
    \psi(-1) \\
    \psi(1)
    \end{matrix}\right)\label{Aeq}\\
    \left(\begin{matrix}
    \psi(-3) \\
    \psi(-1)
    \end{matrix}\right) &= B
    \left(\begin{matrix}
    \psi(-1) \\
    \psi(1)
    \end{matrix}\right)\label{Beq}
\end{align}
With
\begin{align}
    T = \left(\begin{matrix}
        0 & -1 \\
        1 & E
    \end{matrix}\right), &\quad
    S = \left(\begin{matrix}
        E & 1 \\
        -1 & 0
    \end{matrix}\right) \\
    A = \left(\begin{matrix}
        0 & -1 \\
        1+\rho/\lambda & E
    \end{matrix}\right), &\quad
    B = \left(\begin{matrix}
        E & 1+\rho/\lambda \\
        -1 & 0
    \end{matrix}\right)
\end{align}
For a bound state solution to exist, $T$ and $S$ need to have an eigenvalue of absolute value strictly less than 1, and this requires $|E|>2$. 
Furthermore, we need to be able to find a non-zero vector $(\psi(-1),\psi(1))^t\in\Cx^2$ such that \eq{Aeq} and \eq{Beq} yield eigenvectors
$(\psi(1),\psi(3))^t$ and $(\psi(-3),\psi(-1))^t$ belonging to those eigenvalues less than 1 of $T$ and $S$, respectively. Setting $s=1+\rho/\lambda$,
the result of a straightforward calculation leads to the conditions:
\begin{align}
    E &= \pm (s + \frac{1}{s}),\quad |s| >1.
\end{align}
Concretely, this means that for a bound state to exist, one requires (i) $\rho\neq 0$ and (ii) that either $\rho$ and $\lambda$ have the same sign or, if these parameters have opposite signs, 
$|\lambda| < |\rho|/2$.

From numerical results discussed in Section \ref{sec:numerical}, one sees that the most strongly bound states occur, in fact, for $K$ values at the Brillouin zone 
boundary. By standard perturbation theory it is clear that the bound states we found for $K_x=0$ will persist for sufficiently small $|\sin K_x|$. How small is sufficiently small, 
however, may depend on $K_y$.

\section{Connection with the Dirac Equation}\label{sec:dirac}

\begin{figure}
    \centering
    \includegraphics[width=0.49\linewidth]{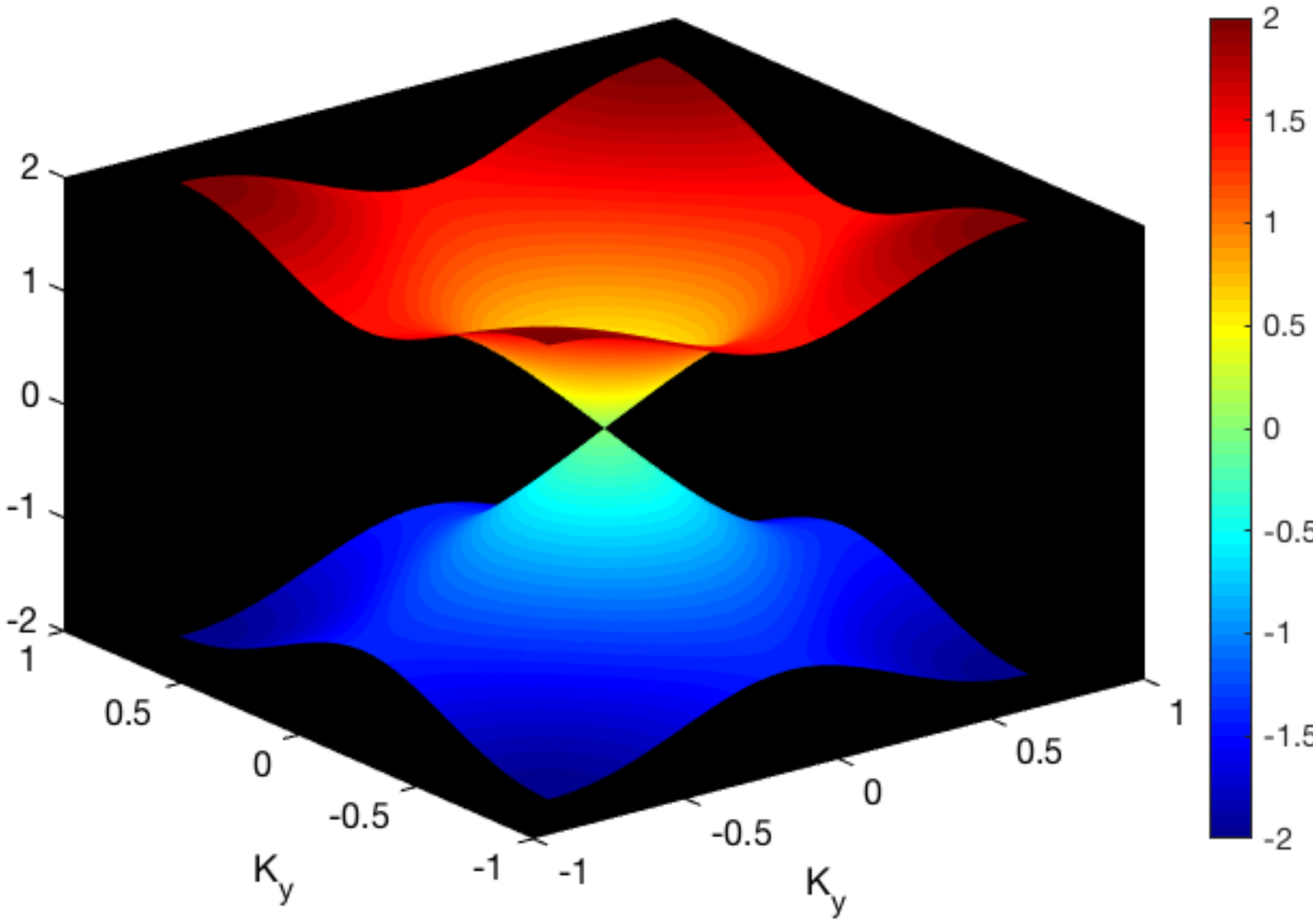}
    \includegraphics[width=0.49\linewidth]{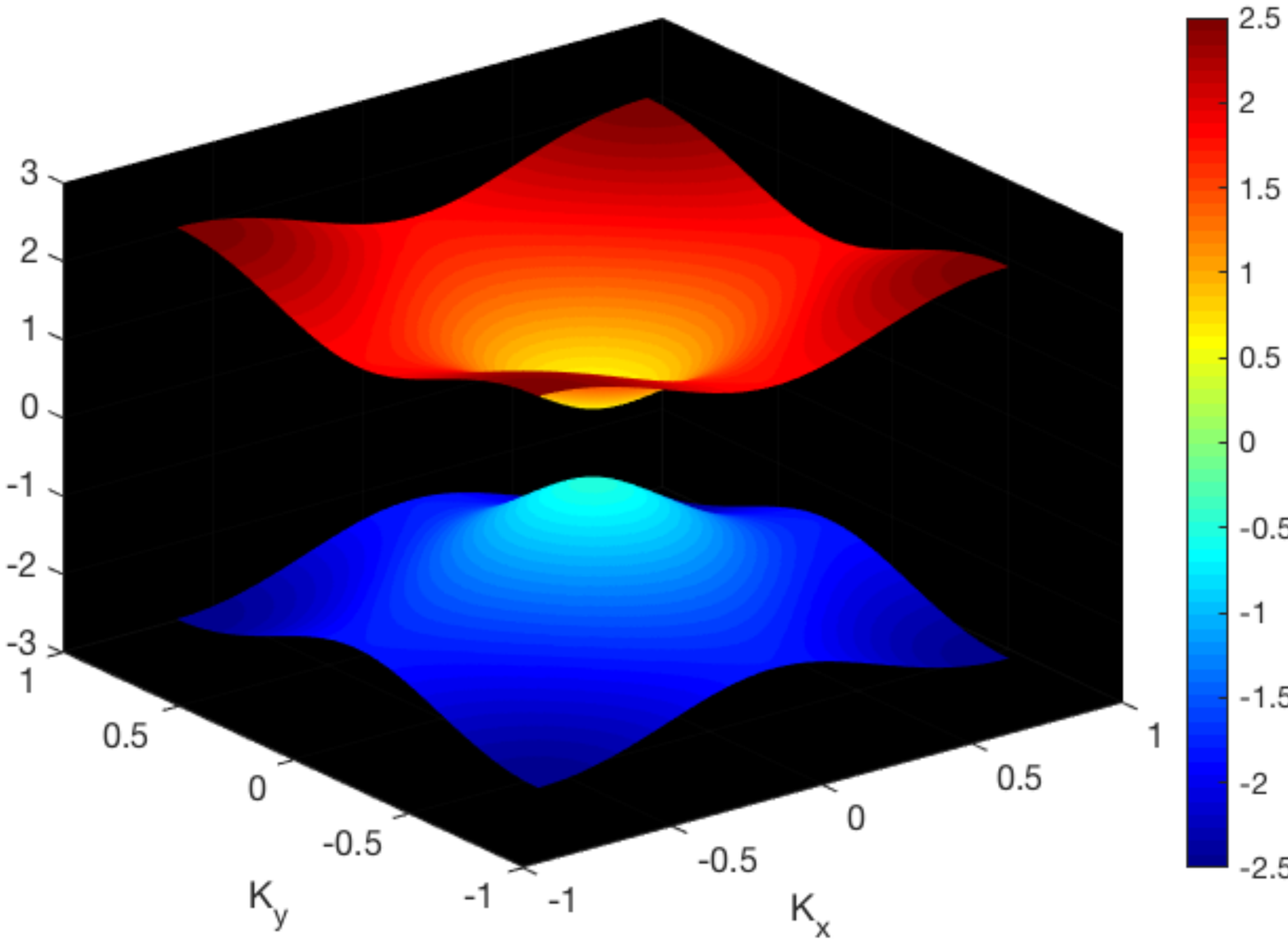}
    \caption{The dispersion relation of the self-dual interaction $H_K^i$ is shown on the left and the one on the right has broken duality with $m=0.5$ as defined in section \ref{sec:dirac}. }
    \label{fig:cone}
\end{figure}

In the small $K$ regime, we in fact have that $H_k^i(K\simeq 0)$ is unitarily equivalent to the massless Dirac Hamiltonian in 2+1 dimensions, analogously to what is observed in graphene. If we break the duality symmetry of $H^{\epsilon\mu}$ with a parameter $m$ by writing
\begin{align}
    H^{\epsilon\mu}=(1+m)&i\sum_{e}\sigma^3_e\sum_{v\in e} s(v,e) A_v\sum_{f\ni e} (\idty - B_f)\\
 + &i\sum_{e}\sigma^1_e\sum_{f\ni e} s(f,e) B_f\sum_{v\in e}(\idty - A_v).
\end{align}
We find the low energy spectrum takes the form 
\begin{equation}
    E(K) \simeq \pm\sqrt{(4K)^2 + 2m^2}
\end{equation}
which is the dispersion relation of a massive relativistic particle, and $H_k^i(K\simeq 0)$ is unitarily equivalent to a massive Dirac Hamiltonian in 2+1 dimensions in this case. In Figure \ref{fig:cone}, we show the dispersion relation of $H_k^i$ in both cases with $m=0$, and $m\neq 0$.

\section{Numerical results}\label{sec:numerical}
We also perform exact diagonalization of the DTCM on a square lattice of length $2L+1$,
with both periodic and free boundary conditions. 
To do this, we simply use the matrix elements $h_{d',d}(K)$ obtained in the thermodynamic limit implemented on a finite lattice. 
In the case of free boundary conditions, we just ignore hopping that would cause a particle to leave the finite lattice. In all numerical results, we fixed $\rho=1$, and $\lambda_{\epsilon} = \lambda_{\mu} := \lambda$. 

\begin{figure}[t]
\includegraphics[width=0.49\columnwidth]{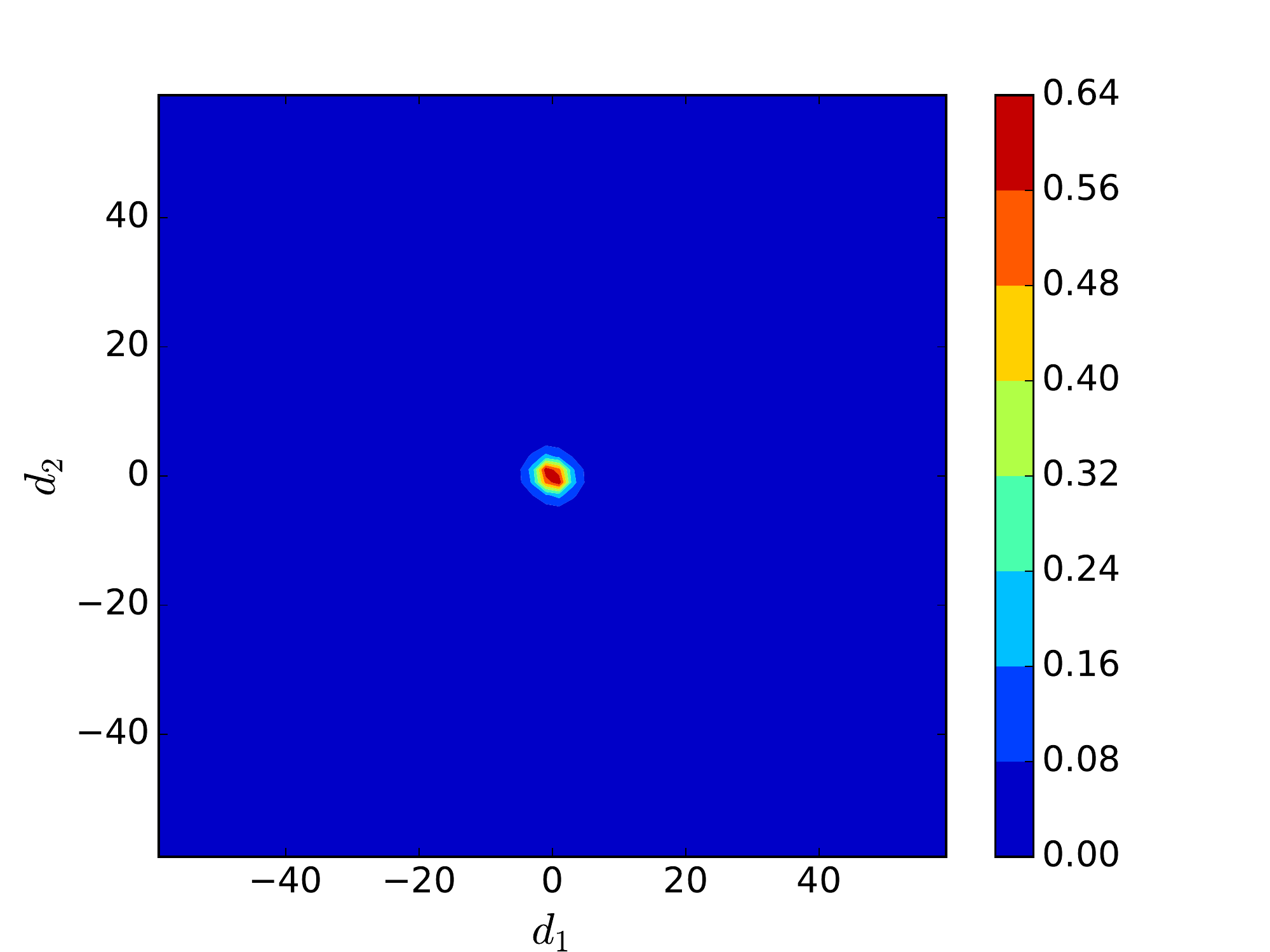}
\includegraphics[width=0.49\columnwidth]{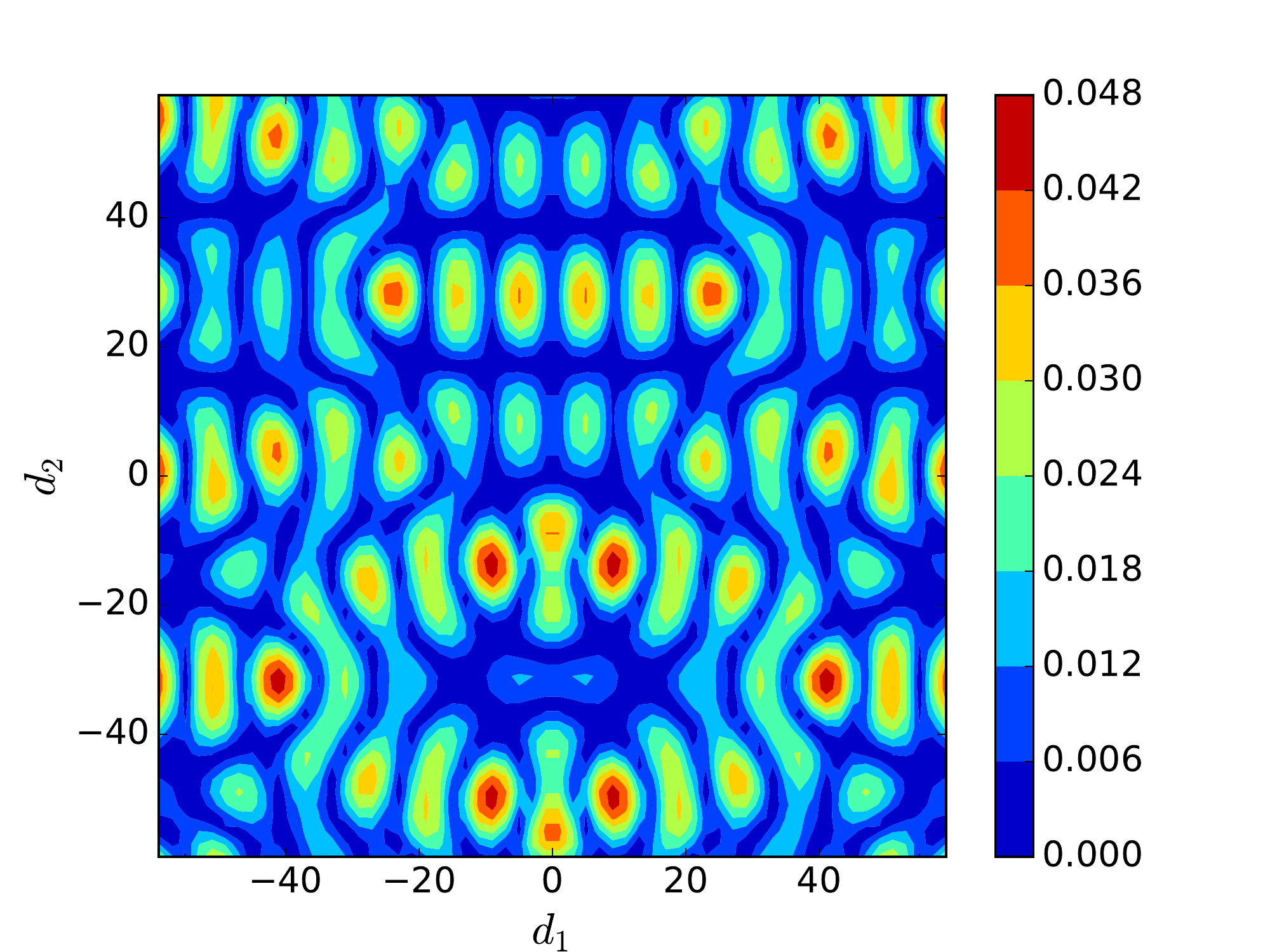}
\caption{Plots of the expansion coefficients of the eigenstates of the DTCM for $L=60$, $\lambda=0.5$, and $K=(\pi/4,\pi/4)$. On the left we show the expansion coefficients for the bound state and on the right we show a typical scattering state found in the middle of the spectrum.}
\label{fig:scattering_state}
\end{figure}

In Figure \ref{fig:scattering_state} we show the absolute value of the expansion coefficients of a bound state and a scattering of this model at $K=(\pi/4,\pi/4)$, $\lambda=0.5$, $L=60$, and using periodic boundary conditions. In the case of the bound state, we see rapid decay of the coefficients with increasing $d$, in agreement with our analysis in the previous section, implying a bound state. We also see in Figure \ref{fig:scattering_state} an example of a scattering state sampled from the middle of the spectrum. Generically the states in the middle of the spectrum take this form, with near equal amplitude for all values of $d$. In both these plots we have fixed $K$, but the qualitative features shown are generally true for all $K$ values that are not $0$.

\begin{figure}[t]
    \centering
    \includegraphics[width=0.7\columnwidth]{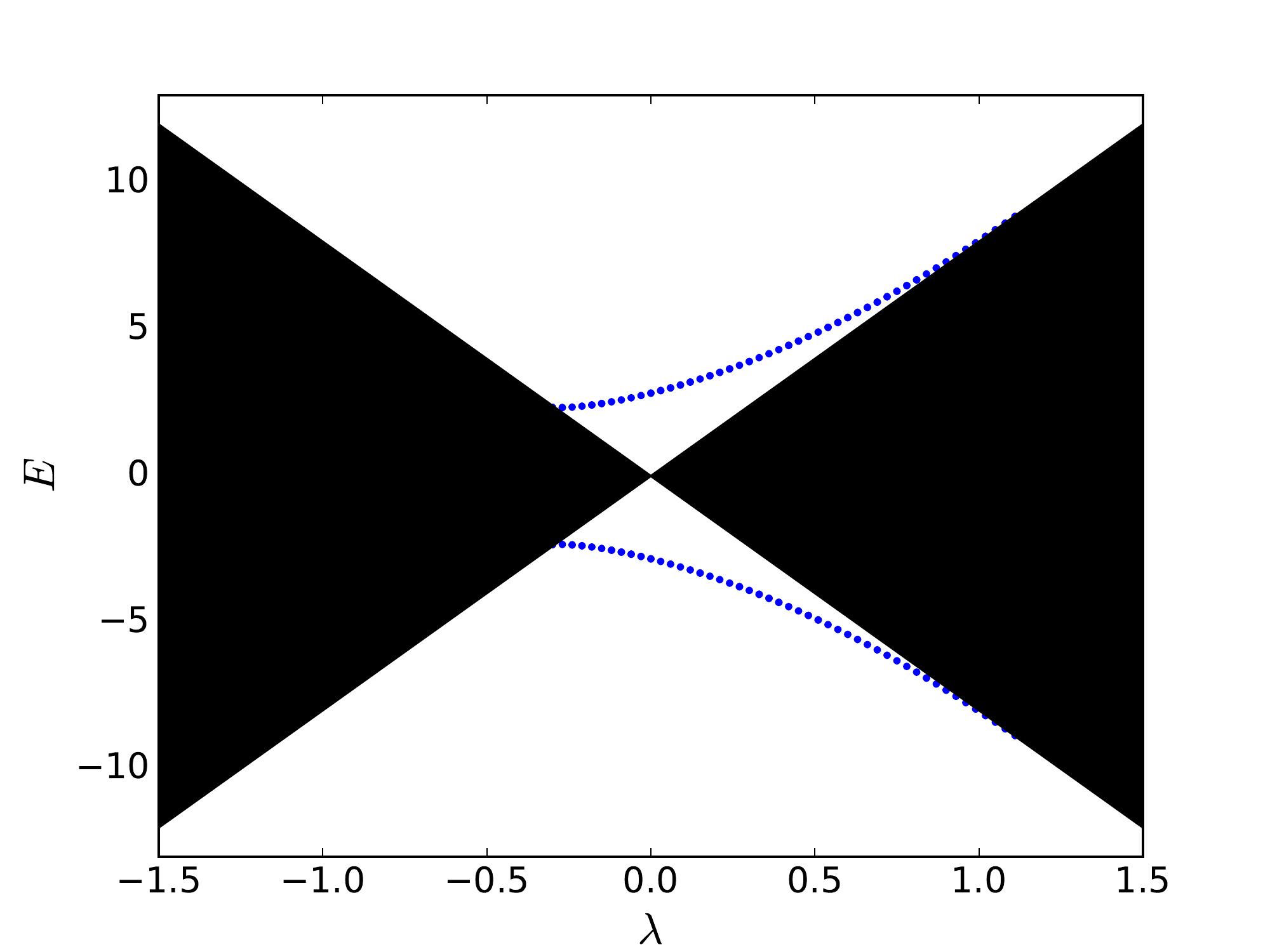}
    \caption{Plot of the spectrum of the DTCM at fixed $K=(\pi/4,\pi/4)$, $\rho=1$, and $L=30$.}
    \label{fig:energies}
\end{figure}

\begin{figure}[t]
    \centering
    \includegraphics[width=0.7\columnwidth]{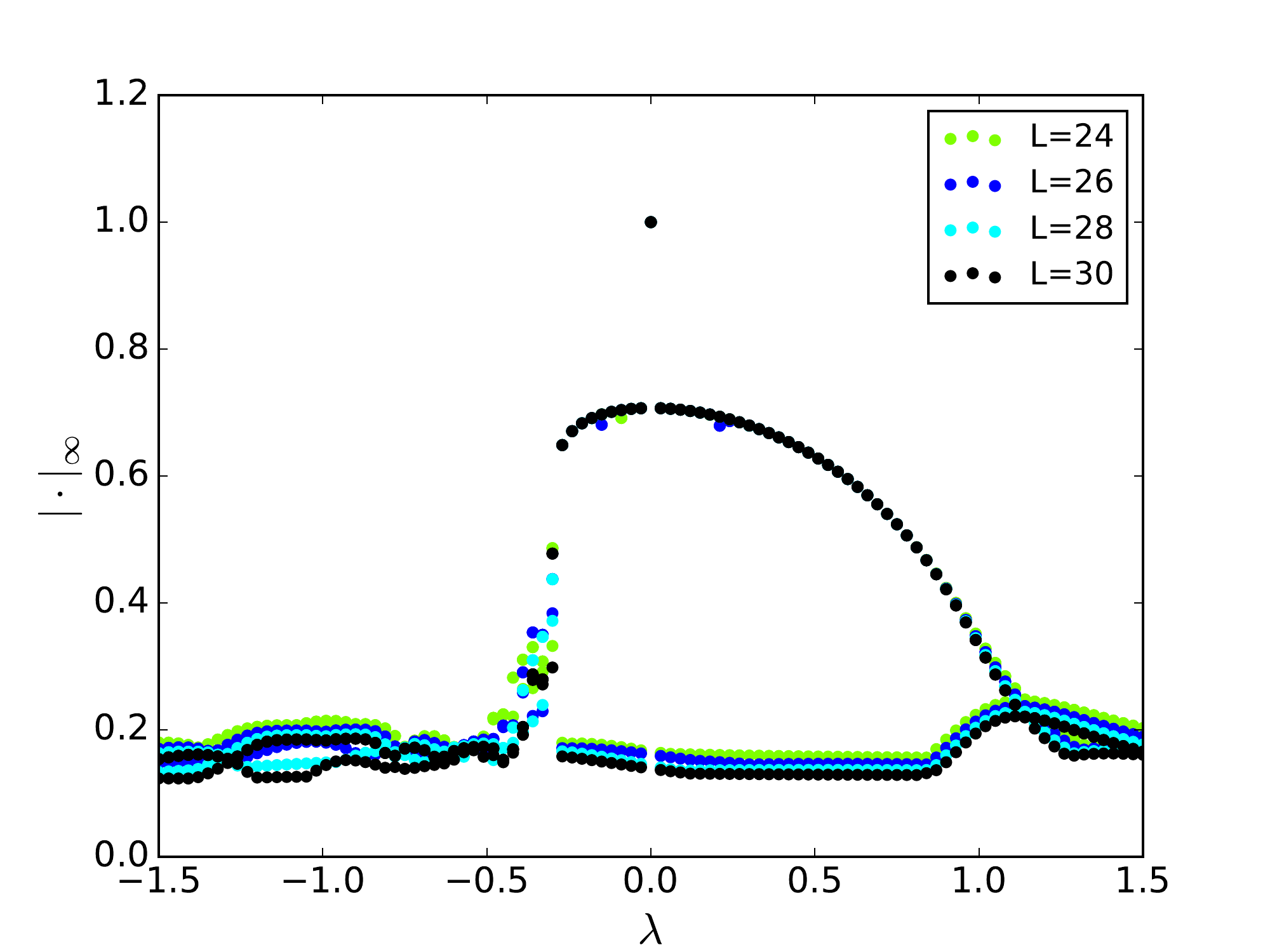}
    \caption{Plot of the $l^{\infty}$ norm of the eigenstates of the DTCM at fixed $K=(\pi/4,\pi/4)$ and $\rho=1$ at various values of the system size $L$.}
    \label{fig:norms}
\end{figure}

In Figure \ref{fig:energies} we show the spectrum of the DTCM for both positive and negative values of $\lambda$, fixed $K=(\pi/4,\pi/4)$, and free boundary conditions. From chirality, the essential spectrum is invariant under $\lambda\rightarrow -\lambda$, but the effect on the bound states is not. When $\lambda=0$, we have the eigenvalues from $H_K^i$, and an extensive degeneracy at $E=0$. As we tune $\lambda$, we see the response of these eigenvalues to the term $H_K^0$. We see that for $\lambda < 0$, we have that the bound state enters the continuum abruptly, while for $\lambda > 0$ the bound states appear to converge to the edges of the continuum band. In Figure \ref{fig:norms} we show the $l^{\infty}$ norm of the two largest distinct eigenvalues of ${h_K}_{d',d}$ as a function of $\lambda$ for varying system sizes. We see that the $l^{\infty}$ norm is robust as we vary system size as expected for a bound state. For $\lambda>0$, the magnitude of the norm decreases in a continuous manner as $\lambda$ increases, but the decrease in norm is abrupt for $\lambda < 0$.

Lastly, we examine the $K$-dependence of the quantity 
\begin{equation}\label{V}
    V(K) = E_{\rm max}(K)-E^0_{\rm max}(K),
\end{equation}
where $E^0_{\rm max}$ is the maximum eigenvalue with $\rho=0$, and $E_{\rm max}$ is the maximum eigenvalue for $\rho=1$, both found numerically at a fixed value of $K$. The quantity $V(K)$ gives a measure of the attraction of the anyons at the given system parameters. If $V(K)$ is positive, it means that the $H_K^i$ term is stronger than the $H_K^0$ term, and so a bound state is expected. If $V(K) \le 0$, then the term $H_K^0$ dominates, and so we expect no bound state.
% Since the essential spectrum of $H^{DTC}$ is that of $H^{\epsilon} + H^{\mu}$, we must have that the total energy has a contribution from the free hopping part given by the first term in $V(K)$. Thus this difference measures the relative strength of the attraction caused by $H^{\epsilon\mu}$, and so $V(K)<0$ implies a bound state. 

\begin{figure}
    \centering
    \includegraphics[width=0.7\linewidth]{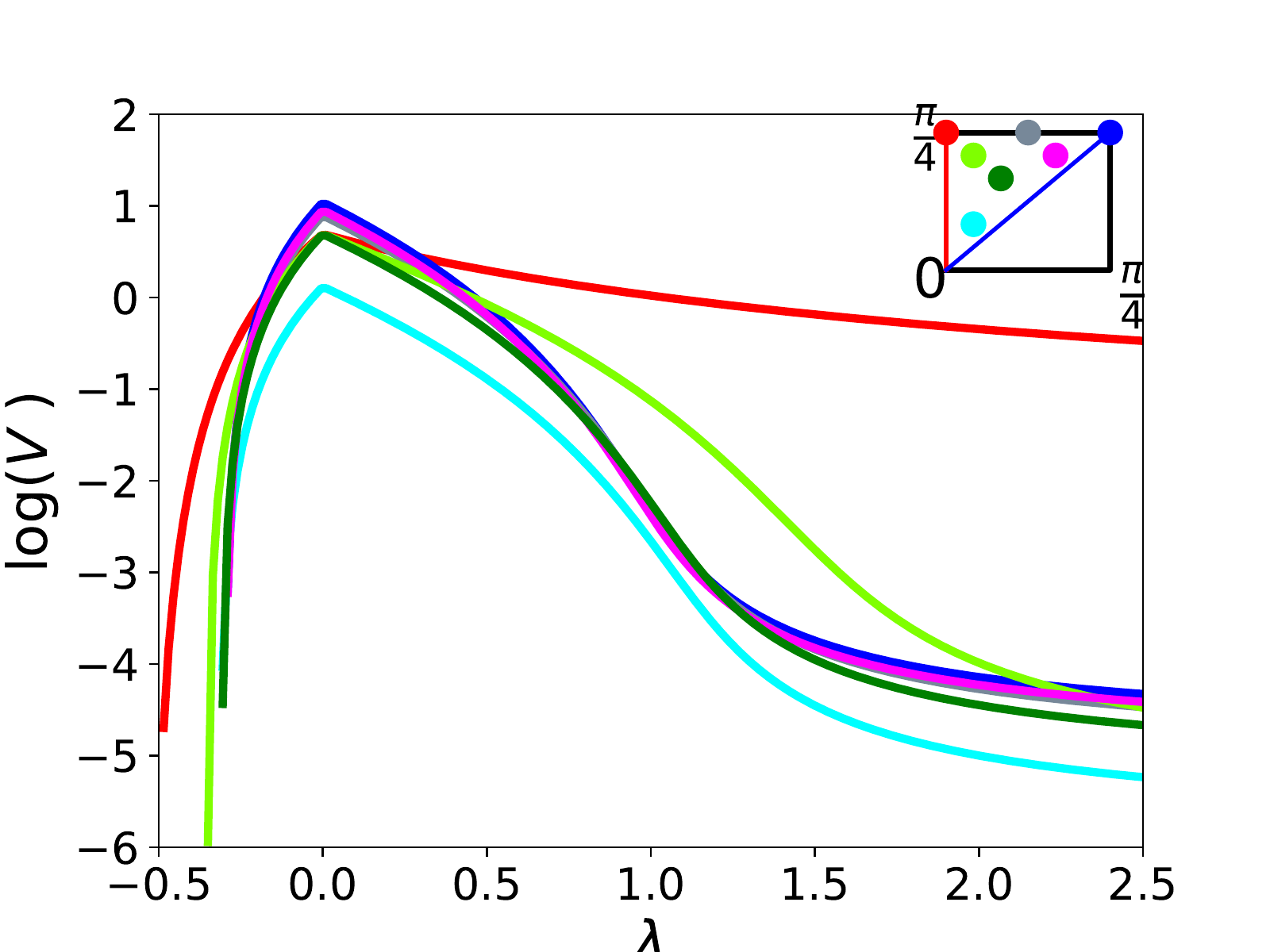}
    \caption{Plot of $V(K)$ defined in Equation \ref{V} as a function of $\lambda$ at various $K$ values, with $\rho=1$, and $L=30$. The inset shows a color key for the $K$ values used. Note that the qualitative behavior is the same for all points on the lines $K_x=K_y$ and $k_x=0$, so we only choose the endpoints. The spectrum is symmetric about the line $K_x=K_y$, so we only show $K_y>K_x$.}
    \label{fig:kdep}
\end{figure}

\begin{figure}
    \centering
    \includegraphics[width=0.7\linewidth]{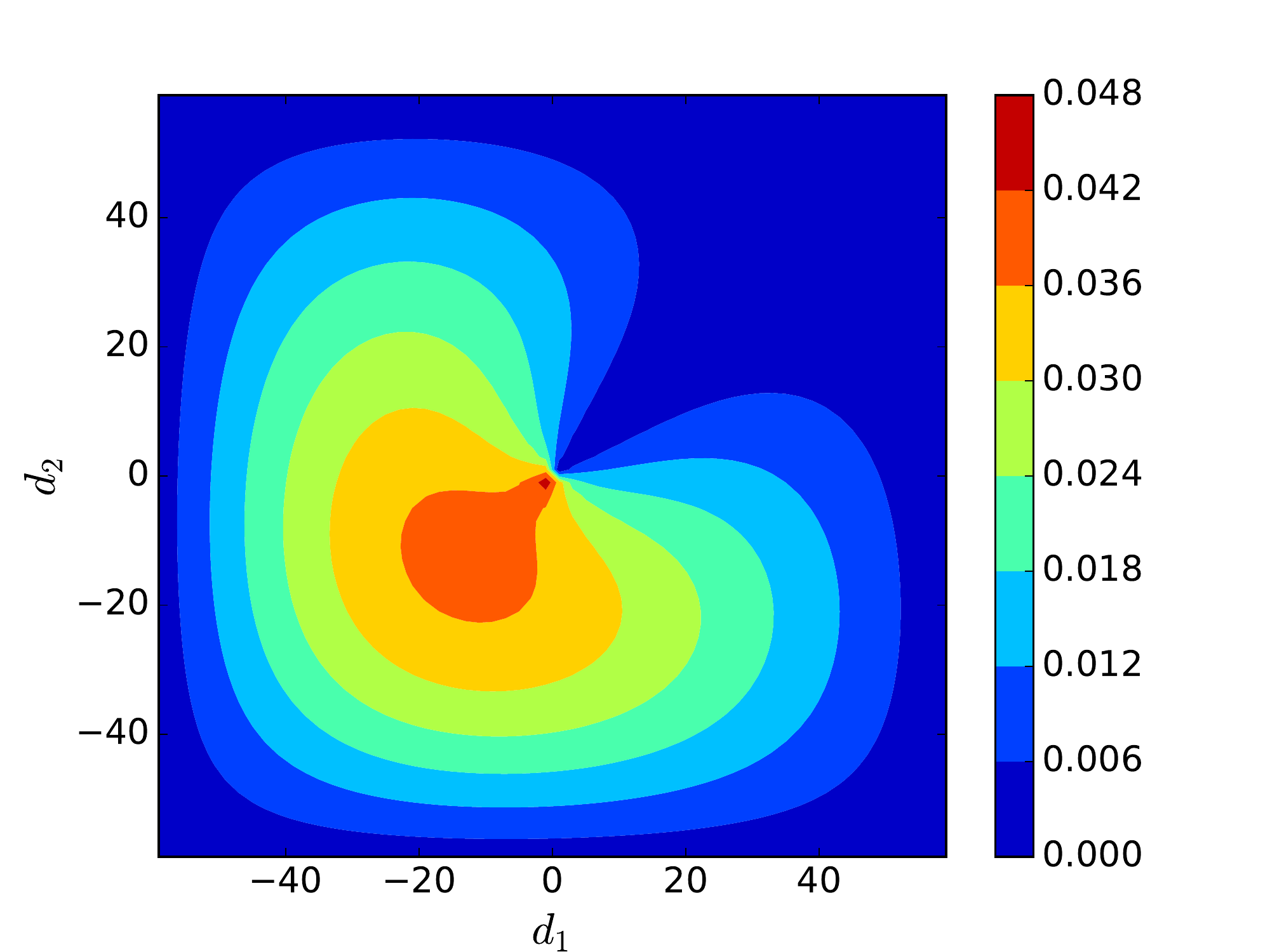}
    \caption{Plots of the expansion coefficients of the maximum eigenstate of the DTCM for $K=(\pi/4,\pi/4)$, $\lambda=1.5$ and $L=60$.}
    \label{fig:rydberg}
\end{figure}

In Figure \ref{fig:kdep} we show $\log V(K)$ for various values of $K$ and $\lambda$, with $\rho=1$, $L=30$, and free boundary conditions. We see that for negative $\lambda$, there is a rapid decay in $V(K)$, and a cutoff where $\log V(K)$ is undefined due to $V(K)$ becoming negative. This suggests that the bound state disappears for negative $\lambda$, in agreement with Figure \ref{fig:energies}, and the analysis in Section \ref{sec:spectrum}. For positive $\lambda$, we see that $V(K)$ remains positive for all $\lambda$ values shown, and in fact $V(K)$ remains positive for $\lambda$ even as large as $\lambda=100$. This suggests that a bound state exists for all $\lambda>0$ for the $K$ values shown. There is a change in the slope of this curve for $\lambda\sim1$, where the states crossover from tightly to loosely bound states. We believe that these $K$ values chosen represent the behavior for all $K$, and that the existence of a bound state for all $\lambda>0$ is true in general for arbitrary $K$. In Figure \ref{fig:rydberg} we show what the maximum eigenstate looks like for $\lambda=1.5$, analogous to Figure \ref{fig:scattering_state}. We see that in this case the expansion coefficients have appreciable magnitude for a large range of $d$ values. We believe this state is loosely bound, similar to Rydberg states in atoms.

% In Figure \ref{fig:kdep} we show $V(K)$ in the first Brillouin zone (BZ) for $L=30$, $\rho=1$, and $\lambda$ \nick{Include lambda values}. We see that $V(K)>0$ for $K$ near the corners of the BZ, suggesting the loss of a bound state solution. We see that for $K\sim 0$ the anyons are only weakly bound, where the strongest attraction is seen in the regime where $K_x\sim 0$ and $K_y$ larger, or vice versa. This picture is true generally for intermediate values of $|\lambda|$. If $\lambda$ is sufficiently small though, we have that $V(K)<0$ for all $K$, and at higher values of $\lambda$ \nick{fill me in}

\section{Discussion}

We introduced a perturbation of Kitaev's Toric Code Hamiltonian that 
turns the static excitations of the TCM into dynamical particles. 
We took care to preserve the essential symmetries of the model.
In particular, the perturbations leave the minimally charged sectors invariant.
We then performed a detailed analysis of the spectrum of the dynamical model 
in the sector charged with one electric and one magnetic anyon. We found that
the `ribbon states' in a certain range of the center of mass momentum are stable,
i.e., exist as a bound state of one electric and one magnetic charge. At a critical value
of the ratio of the parameters $\rho$ and $\lambda$ in the DTCM, the bound state
eigenvalue dips into the band of scattering states, becomes unstable and the electric and magnetic 
anyons de-fuse into separate electric and magnetic charges.

Similar considerations can be applied to the general class of quantum double models
introduced by Kitaev and other constructions of commuting Hamiltonians describing anyons.

\begin{acknowledgments}
BN acknowledges stimulating discussions with Sven Bachmann and Yosi Avron. NS acknowledges helpful discussions with Tomohiro Soejima. 
Based on work supported by the National Science Foundation under grant DMS-1813149 (BN).
\end{acknowledgments}

%\bibliographystyle{amsplain}
%\providecommand{\bysame}{\leavevmode\hbox to3em{\hrulefill}\thinspace}
%\bibliography{qss}
\providecommand{\bysame}{\leavevmode\hbox to3em{\hrulefill}\thinspace}
\providecommand{\MR}{\relax\ifhmode\unskip\space\fi MR }
\providecommand{\MRhref}[2]{%
  \href{http://www.ams.org/mathscinet-getitem?mr=#1}{#2}
}
\providecommand{\href}[2]{#2}

 %\bibitem{footnote1} In fact, there is complex, non translation-invariant gauge transformation that makes the hopping matrix elements real, but working in that gauge would not offer any advantages.
%\end{thebibliography} 
\end{document}